\documentclass[aps,showpacs,preprint,preprintnumbers,amsmath,amssymb,nofootinbib]{revtex4}
\usepackage{epsfig}
\usepackage{graphicx}
\usepackage{dcolumn}
\usepackage{bm}
\usepackage{feynmp}
\usepackage{slashed}

\def\beq{\begin{equation}}
\def\eeq{\end{equation}}

\newcommand{\bea}{\begin{eqnarray}}
\newcommand{\eea}{\end{eqnarray}}
\def\Larg{\mathcal L}

\def\eeqn{\end{equation}}
\newcommand\iden{\leavevmode\hbox{\small1\normalsize\kern-.33em1}}

\let\jnfont=\rm
\def\NPB#1,{{\jnfont Nucl.\ Phys.\ B }{\bf #1},}
\def\PLB#1,{{\jnfont Phys.\ Lett.\ B }{\bf #1},}
\def\EPJC#1,{{\jnfont Eur.\ Phys.\ Jour.\ C }{\bf #1},}
\def\PRD#1,{{\jnfont Phys.\ Rev.\ D }{\bf #1},}
\def\PRL#1,{{\jnfont Phys.\ Rev.\ Lett.\ }{\bf #1},}
\def\MPLA#1,{{\jnfont Mod.\ Phys.\ Lett.\ A }{\bf #1},}
\def\JPG#1,{{\jnfont J.\ Phys.\ G }{\bf #1},}
\def\CTP#1,{{\jnfont Commun.\ Theor.\ Phys.\ }{\bf #1},}
\def\JHEP#1,{{\jnfont JHEP \ }{\bf #1},}
\def\NPPS#1,{{\jnfont Nucl.\ Phys.\ Proc.\ Suppl.\ }{\bf #1},}

\begin{document}


\title{Supersymmetric Extension of the Minimal Dark Matter Model}

\author{{Xue Chang}$^1$, {Chun Liu}$^1$, {Feng-Cai Ma}$^2$, {Shuo Yang}$^{3,4}$}
\affiliation{
$^1$ State Key Laboratory of Theoretical Physics,
Institute of Theoretical Physics, Chinese Academy of Sciences,
P.O. Box 2735, Beijing 100190, China\\
$^2$ Physics Department, Liaoning University ,
Shenyang, 110036, P.R China\\
$^3$ Physics Department, Dalian University,
Dalian, 116622, P.R China\\
$^4$ Center for High-Energy Physics, Peking University,
Beijing, 100871, P.R China}
\email{chxue@itp.ac.cn, liuc@mail.itp.ac.cn, fcma@lnu.edu.ca, yangshuo@dlu.edu.cn}

\begin{abstract}

The minimal dark matter model is given a supersymmetric extension.  A
super $SU(2)_L$ quintuplet is introduced with its fermionic neutral
component still being the dark matter, the dark matter particle mass is about 
19.7 GeV. Mass splitting among the quintplet due to supersymmetry particles
is found to be negligibly small compared to the electroweak
corrections. Other properties of this supersymmetry model are studied, 
it has the solutions to the PAMELA and Fermi-LAT anomaly, the predictions 
in higher energies need further experimental data to verify.

\end{abstract}


\maketitle

\section{Introduction}

It has been known from many astrophysical measurements that the
universe contains enormous of invisible, non-baryonic dark matter (DM)
which is not included in the Standard Model (SM).  Among various
hypotheses for the nature of the DM, that of weakly interacting massive
particles (WIMPs) is very attractive. Focussing on the DM problem, one
can explore a simple WIMP model, that is the minimal dark matter model
(MDM) \cite{mdm1,mdm2}: adding to the SM a single matter $\mathrm{X}$ without
introducing any additional discrete symmetry, and $\mathrm{X}$ is in a high
dimensional representation of the  usual SM $SU(2)_L\times U(1)_Y$
electroweak (EW) symmetry. The stability of the DM candidate is
guaranteed by the SM gauge symmetry and by the renormalizability.  The
minimality of the model lies in the fact that the new physics is
determined by only one parameter, namely the mass $M$ of the
$\mathrm{X}$ multiplet. Therefore, the MDM is remarkably predictive.
There are some extensions to the MDM \cite{others}.

As far as the particle physics is concerned, the SM provides a
successfully description of presently known phenomena. However it will
have to be extended to describe physics at higher energies. That often
results in the gauge hierarchy problem. Supersymmetry (SUSY) offers a
solution to this problem \cite{susy}.

In this work, we make SUSY extension to the MDM. Note that in the
so-called minimal SUSY extension of the SM (MSSM), the DM candidate,
that is the lightest SUSY particle, is there only after introducing an
extra discrete symmetry by hand, which is the $R$-parity. In our SUSY
MDM (SMDM), instead, we still follow the logic of MDM, the existence of
the DM lies in the fact that the DM is in a high dimensional
representation of the SM gauge group without using discrete symmetries.

In section II, the SMDM is constructed.  In section III, mass splitting
of the $\mathrm{X}$ multiplet, the DM relic density, direct and
indirect detection signatures of the SMDM are calculated.  In section  IV,
the conclusion is made. In the Appendix, we give basic facts about the
representation of the SU(2) group.

\section{SMDM}

The SMDM is simply constructed by supersymmetric extension to the MDM. The
particle content is, in addition to that of the MSSM, the fermionic
$SU(2)_L$ 5-plet $\mathrm{X}$ of the MDM and its superpartner which a
complex scalar 5-plet $\tilde{\mathrm{X}}$. The charged components
$\mathrm{X}^{Q}$ are slightly heavier than the neutral one
$\mathrm{X}^0$ due to quantum corrections \cite{mdm1,mdm2}; and the superpartner
$\tilde{\mathrm{X}}$ because of its soft mass $M_{soft}$, is also
heavier than $\mathrm{X}$.  Both $\mathrm{X}^{Q}$ and
$\tilde{\mathrm{X}}$ will decay into $\mathrm{X}^0$. The relic particle in
the SMDM is still $\mathrm{X}^0$ as in the MDM. The new
parameters are $M$ and $M_{soft}$, the model is still predictive.
As for the Lagrangian, in addition to that of the MSSM, we have
\begin{equation}
\label{smdm}
\begin{aligned}
\Larg_{SMDM}&=
   \frac{i}{2}
   (\mathrm{X}^\dagger {^i }\bar\sigma^\mu\partial_\mu\mathrm{X}_i
   + \bar{\mathrm{X}}^\dagger{^i }\bar\sigma^\mu
   \partial_\mu\bar{\mathrm{X}}_i)
   - \frac{i}{2} g_2 A_a^\mu(\mathrm{X}^\dagger {^i }\bar\sigma_\mu
   (T^a)_i{^j}\mathrm{X}_j
   -\bar{\mathrm{X}}^\dagger {_i }\bar\sigma_\mu(T^a)_j{^i}
   \bar{\mathrm{X}}_j)\\
  &- \frac{\sqrt{2}}{2} g_2 (\tilde{\mathrm{X}}^*{^i}
   (T^a)_i{^j}\mathrm{X}_j\lambda^{a}
   + \lambda^\dagger {^a }\mathrm{X}^\dagger {^i }
   (T^a)_i{^j}\tilde{\mathrm{X}}_j
   -\tilde{\bar{\mathrm{X}}}^*{^i}(T^a)_j{^i}\bar{\mathrm{X}}_j
   \lambda^{a}
   -\lambda^\dagger {^a }\bar{\mathrm{X}}^\dagger {^i }
   (T^a)_j{^i}\tilde{\bar{\mathrm{X}}}_j)\\
  &+ \frac{1}{2}(D^\mu \tilde{\mathrm{X}}^*D_\mu\tilde{\mathrm{X}}
   -M^2 |\tilde{\mathrm{X}}|^2)
   -\frac{1}{2}M(\bar{\mathrm{X}}_i \mathrm{X}_i
   + \mathrm{X}^\dagger {^i }\bar{\mathrm{X}}^\dagger {^i })
   +\frac{1}{2}g_2 D_a {\tilde{\mathrm{X}}}^i
   (T^a)_i{^j} {\tilde{\mathrm{X}}}_j\\
  & -\frac{1}{2}M_{soft}^2 |\tilde{\mathrm{X}}|^2.
\end{aligned}
\end{equation}
The component field notation has been used.  In eq.(1), $T^a$'s  are
generators of the $SU(2)_L$ in $\bold{n}$ representation. $\mathrm{X}_i$
and $\bar{\mathrm{X}}^i$ consist of the left-hand pairs of $\mathrm{X}_i$,
transforming in $SU(2)_L$ $\bold{5}$ representation with the generator
$(T^a)_i{^j}$ and the complex conjugate representation with the generator
$(T^a)^*_i{^j}=(T^a)_j{^i}$, respectively. They are not independent.  Actually they are
dual to each other under the $SU(2)_L$.  We write the Lagrangian in the
form of eq.(\ref{smdm}) just for convenience.  Their superpartners compose
a bosonic $SU(2)_L$ 5-plets $\tilde{\mathrm{X}}^i$.  Both $\mathrm{X}_i$
and $\tilde{\mathrm{X}}^i$ have trival $SU(3)_c\times U(1)_Y$ quantum
numbers $(1,0)$.

In terms of 4-component notation, we can define following spinors,
\begin{equation}
   \Psi^{+2}\equiv
\begin{pmatrix}
   \mathrm{X}^{+2}
  \\ (\bar{\mathrm{X}}^{+2})^\dagger
\end{pmatrix},
   \Psi^{-2}\equiv
\begin{pmatrix}
   \mathrm{X}^{-2}
  \\ (\bar{\mathrm{X}}^{-2})^\dagger
\end{pmatrix}
   =(\Psi^{+2}){^C},
\end{equation}

\begin{equation}
   \Psi^{+1}\equiv
\begin{pmatrix}
   \mathrm{X}^{+1}
  \\ (\bar{\mathrm{X}}^{+1})^\dagger
\end{pmatrix},
   \Psi^{-1}\equiv
\begin{pmatrix}
   \mathrm{X}^{-1}
  \\ (\bar{\mathrm{X}}^{-1})^\dagger
\end{pmatrix}
   =(\Psi^{+1}){^C},
\end{equation}

\begin{equation}
   \Psi^{0}\equiv
\begin{pmatrix}
   \mathrm{X}^{0}
  \\ (\bar{\mathrm{X}}^{0})^\dagger
\end{pmatrix}
   =(\Psi^{0}){^C}.
\end{equation}
The neutral field $\Psi^{0}$ is a Majorana field.

The superpotential takes the simple from:
\begin{equation}
   W=W_{MSSM}+\frac{1}{2}M\tilde{\mathrm{X}}^2,
\end{equation}
which gives $\mathrm{X}$ and $\tilde{\mathrm{X}}$ the same unbroken
supersymmetry mass $M$. $\tilde{\mathrm{X}}$ get a soft mass
after supersymmetry breaking. The $D$-term contribution to the scalar
potential is:

\begin{equation}
   V_D=\frac{1}{2}
   g_2^2\left[\sum\phi^* t^a\phi+
   \frac{1}{2}\tilde{\mathrm{X}}^*T^a\tilde{\mathrm{X}}\right]^2.
\end{equation}
where $\phi$ denotes the $SU(2)_L$ scalars in the MSSM. Compared with
the MSSM, the extra term is:
\begin{equation}
   \frac{1}{2} g_2^2\left[\sum(\phi^* t^a \phi)
   (\tilde{\mathrm{X}}^* T^a \tilde{\mathrm{X}})
   +\left(\frac{1}{2}\tilde{\mathrm{X}}^* T^a
   \tilde{\mathrm{X}} \right)^2\right].
\end{equation}
These couplings do not cause $\tilde{\mathrm{X}}$ dacay but annihilation
into MSSM $SU(2)_L$ scalars, which give an extra negligible mass
splitting between $\tilde{\mathrm{X}}^i$.

Considering non-renormalizable terms of the lagrangian, there are dimension
5 operators $\tilde{\mathrm{X}}^{ijkl} \phi_i\phi_j\phi_k \phi_l/\Lambda$ for
the complex scalar 5-plet $\tilde{\mathrm{X}}$, and dimension 6 operators
$\mathrm{X}^{ijkl} \psi_i\phi_j\phi_k \phi_l /\Lambda^2$ for the
fermionic 5-plet $\mathrm{X}$ allowed by the $SU(2)_L\times U(1)_Y$ gauge
symmetry, where $\psi_i$ is the left-hand leptons or the higgsinos in the
MSSM, eg.
$\tilde{\mathrm{X}}H_uH_dH_uH_u^*/\Lambda$,
$\tilde{\mathrm{X}}H_uH_dH_dH_d^*/\Lambda$,
$\mathrm{X}\tilde{H_u}H_dH_dH_d^*/\Lambda^2$,
$\mathrm{X}LH_uH_dH_u/\Lambda^2$\ ,etc.

We can generate these couplings by adding the corresponding higher
dimension superpotential, eg.
\begin{equation}
W_{non-ren}=\frac{\tilde{\mathrm{X}} H_uH_dH_uH_d}{\Lambda^2}
+ \frac{\tilde{\mathrm{X}} \tilde{L}H_uH_dH_u}{\Lambda^2}+...\ ,
\end{equation}
the equation of motion for the auxiliary fields are:
\begin{equation}
\begin{aligned}
  &F_{H_d}=-(\frac{\partial W}{\partial H_d})^*=
   -(\mu H_u+\frac{\tilde{\mathrm{X}} H_uH_uH_d}{\Lambda^2}+
   \frac{\tilde{\mathrm{X}} \tilde{L}H_uH_u}{\Lambda^2}+...)^*,\\
  &F_{H_u}=-(\frac{\partial W}{\partial H_u})^*=
   -(\mu H_d+\frac{\tilde{\mathrm{X}} H_uH_dH_d}{\Lambda^2}+
   \frac{\tilde{\mathrm{X}} \tilde{L}H_uH_d}{\Lambda^2}+...)^*.
\end{aligned}
\end{equation}
This generates dim6 couplings for the fermionic 5-plet :
$\mathrm{X}\tilde{H_u}H_dH_dH_d^*/\Lambda^2$,
$\mathrm{X}L H_uH_dH_u/\Lambda^2$ where $\Lambda\approx10^{15}$GeV.
These operators can induce 4-body decays with a typical life-time
$\tau \sim {\Lambda^4\, {\rm TeV}^{-5}}\sim10^{19}\rm{s}$ which is
longer than the age of the universe ($\sim10^{17}\rm{s}$).
So these couplings have no influence on the observed stability of the
DM candidates. The F-term also generates the scalar dim.5 operators
but which are not suppressed by one power of $\Lambda$ but two powers:
$\mu\tilde{\mathrm{X}}H_uH_dH_dH_d^*/\Lambda^2$,
$\mu\tilde{\mathrm{X}}\tilde{L}H_dH_dH_d^*/\Lambda^2$.
The typical life-time of these operators is:
$\tau \sim {\Lambda^3\, {\rm TeV}^{-3}\mu^{-1}}\sim10^{20}\rm{s}$ which
is long enough. Of course there are even higher dimension couplings of the
complex scalar
$\tilde{\mathrm{X}}$, eg.
$\tilde{\mathrm{X}}\tilde{\mathrm{X}}^*H_uH_u^*H_dH_d^*H_dH_d^*/\Lambda^4$,
$\tilde{\mathrm{X}}\tilde{\mathrm{X}}^*\tilde{L}H_uH_dH_u^*H_d^*H_d^*/\Lambda^4$,
etc. These higher dimension operators can be neglected in considering the decay.

Therefore the new introduced particles
$(\tilde{\mathrm{X}}^{\pm2},\tilde{\mathrm{X}}^{\pm1},\tilde{\mathrm{X}}^0)$
only decay into $(\mathrm{X}^{\pm2},\mathrm{X}^{\pm1},\mathrm{X}^0)$ via
gauuge interactions.  $(\mathrm{X}^{\pm2},\mathrm{X}^{\pm1},\mathrm{X}^0)$
are quite stable.  We will further study mass splitting among them in the next
section, and see that the DM candidate is still $\mathrm{X}^0$.

\section{Properties of the SMDM}

The DM candidate in the SMDM model is still $\mathrm{X}^0$. Mass splitting
due to SUSY particles is small because of SUSY breaking.

\subsection{Mass splitting}

Mass splitting should be studied in detail, like that in the MDM,
because it can be calculated with little uncertainties in this simple
model. Because of EW symmetry breaking, the gauge kinetic terms gives
the fermonic 5-plet $\mathrm{X}$ a mass splitting through loop
crrections \cite{mdm1,mdm2},
$\Delta M^Q_{EW}\equiv M^Q-M^0\approx Q^2\times 166$ MeV,
where $M{^Q}$ and $M{^0}$ are the pole masses of $\mathrm{X}^Q$ and
$\mathrm{X}^0$, respectively.

The scalar particles $\tilde{\mathrm{X}}$'s also contribute to mass splitting
of $\mathrm{X}$. They are heavier than the fermions $X$'s by a soft mass
$M_{soft}$ which is generally expected to be about 100 GeV-1 TeV. This further
mass splitting is calculated by using the supersymmetric kinetic term,
the second line of the eq.(1), at the loop level which involves
$\tilde{\mathrm{X}}$'s and the gauginos. By using the two-component notation for
fermions, the one-loop pole mass is written as \cite{two-component}
\begin{equation}
   M{^Q_{SUSY}}=M(1+\frac{1}{2}\Sigma_L{^Q}+\frac{1}{2}\Sigma_R{^Q})
\end{equation}
where $\Sigma_L{^Q}$,$\Sigma_R{^Q}$ are the 1PI self-energy functions as shown
in Fig.1 in $Q=0,+1$ cases.

\vspace{1pt}
\begin{figure}[htbp]
\centering
\includegraphics[width=6in]{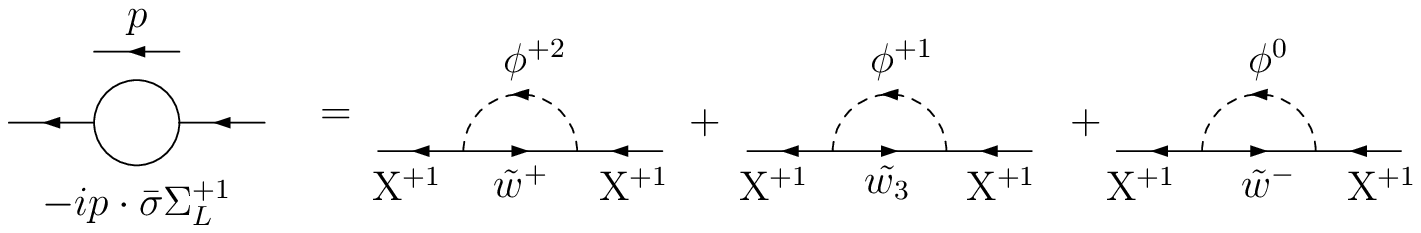}
\includegraphics[width=6in]{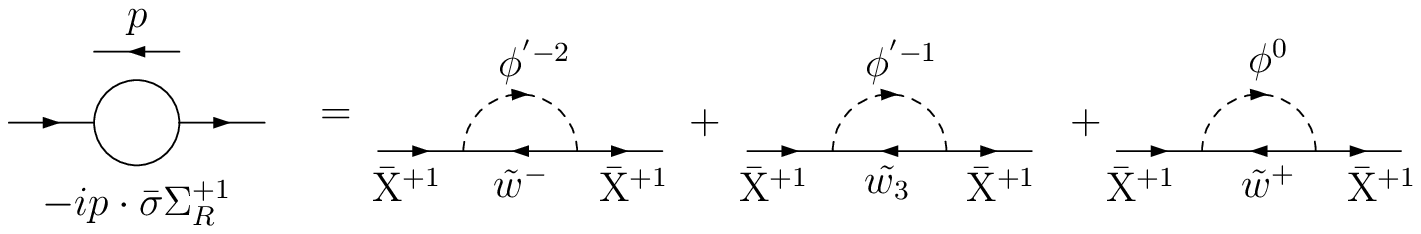}
\includegraphics[width=4.5in]{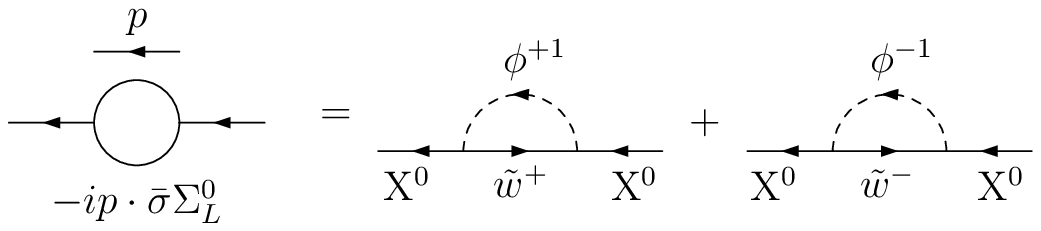}
\includegraphics[width=4.5in]{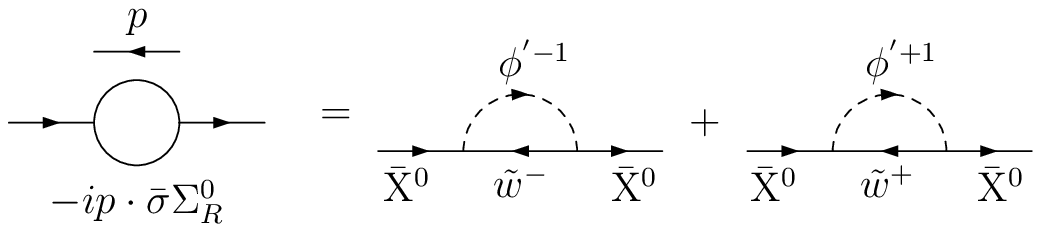}
\caption{One-loop corrections to the 1PI self-energy functions
to the $Q=0,+1$ components of the SMDM.}
\end{figure}
In the diagrams of Fig.1, we denote the correspondent superpartners :
\begin{equation}
   \phi^{+2}\equiv\tilde{\mathrm{X}}^{+2},\phi^{'-2}\equiv\tilde{\bar{\mathrm{X}}}^{+2},
   \phi^{+1}\equiv\tilde{\mathrm{X}}^{+1},\phi^{'-1}\equiv\tilde{\bar{\mathrm{X}}}^{+1},
   \phi^{0}\equiv\tilde{\mathrm{X}}^{+1}.\\
\end{equation}
$\tilde{w}^\pm$, $\tilde{w}_3$ are the
superpartners of the $SU(2)_L$ gauge bosons
and  $\tilde{w}^+=V_{11}^{-1}\tilde{C}_1^{+}+V_{12}^{-1}\tilde{C}_2^{+}$,
$\tilde{w}^-=U_{11}^{-1}\tilde{C}_1^{-}+U_{12}^{-1}\tilde{C}_2^{-}$,
$\tilde{w}_3=N_{2i}^{-1}\tilde{N}_i$, $i$=1-4. $\tilde{C}_1^{\pm}$,
$\tilde{C}_2^{\pm}$ and $\tilde{N}_i$ are the charginos and neutralinos of the MSSM.
$U, V, N$ are the unitary matrices diagonalizing the mass matrices of charginos and
neutralinos \cite{LSP}.

It is worthy to note that the $\Sigma_D{^Q}$ term which is related to the
B-term
$B^{ij}\tilde{\mathrm{X}_i}\tilde{\mathrm{X}_j}$, may also appear in the
pole mass formula, and it does not cause divergences.
But in our calculation we do not consider it for simplicity, it
is enough for us to break the supersymmetry only through the soft mass $M_{soft}$.

Using the superpartner notation mentioned in Sect. I we get :
\begin{equation}
\begin{aligned}
   \Sigma_L^{+1}&=\frac{g_2^2}{16\pi^2}[V^*_{11}V_{11}B{_1}(\tilde{C_1},\phi^{+2})
   +V^*_{21}V_{21}B{_1}(\tilde{C_2},\phi^{+2})
   +\frac{3}{2}(U^*_{11}U_{11}B{_1}(\tilde{C_1},\phi^{0})\\
  &+U^*_{21}U_{21}B{_1}(\tilde{C_2},\phi^{0}))
   +\frac{1}{2}(N^*_{12}N_{12}B{_1}(\tilde{N_1},\phi^{+1})+N^*_{22}N_{22}B{_1}(\tilde{N_2},\phi^{+1})\\
  &+N^*_{32}N_{32}B{_1}(\tilde{N_3},\phi^{+1})+N^*_{42}N_{42}B{_1}(\tilde{N_4},\phi^{+1})],
\end{aligned}
\end{equation}

\begin{equation}
\begin{aligned}
   \Sigma_R^{+1}&=\frac{g_2^2}{16\pi^2}[U^*_{11}U_{11}B{_1}(\tilde{C_1},\phi^{'-2})
   +U^*_{21}U_{21}B{_1}(\tilde{C_2},\phi^{'-2})
   +\frac{3}{2}(V^*_{11}V_{11}B{_1}(\tilde{C_1},\phi^{0})\\
  &+V^*_{21}V_{21}B{_1}(\tilde{C_2},\phi^{0}))
   +\frac{1}{2}(N^*_{12}N_{12}B{_1}(\tilde{N_1},\phi^{'-1})+N^*_{22}N_{22}B{_1}(\tilde{N_2},\phi^{'-1})\\
  &+N^*_{32}N_{32}B{_1}(\tilde{N_3},\phi^{'-1})+N^*_{42}N_{42}B{_1}(\tilde{N_4},\phi^{'-1})],
\end{aligned}
\end{equation}

\begin{equation}
\begin{aligned}
   \Sigma_L^{0}&=\frac{g_2^2}{16\pi^2}[\frac{3}{2}(V^*_{11}V_{11}B{_1}(\tilde{C_1},\phi^{+1})
   +V^*_{21}V_{21}B{_1}(\tilde{C_2},\phi^{+1})
   +U^*_{11}U_{11}B{_1}(\tilde{C_1},\phi^{-1})\\
  &+U^*_{21}U_{21}B{_1}(\tilde{C_2},\phi^{-1}))],
\end{aligned}
\end{equation}

\begin{equation}
\begin{aligned}
   \Sigma_R^{0}&=\frac{g_2^2}{16\pi^2}[\frac{3}{2}(U^*_{11}U_{11}B{_1}(\tilde{C_1},\phi^{'-1})
   +U^*_{21}U_{21}B{_1}(\tilde{C_2},\phi^{'-1})
   +V^*_{11}V_{11}B{_1}(\tilde{C_1},\phi^{'+1})\\
  &+V^*_{21}V_{21}B{_1}(\tilde{C_2},\phi^{'+1}))],
\end{aligned}
\end{equation}
where $B_1$ is the one rank two point integral
\begin{equation}
B{_1}(p^2,m_1,m_2)=
-\frac{1}{2\varepsilon}+\frac{A{_0}(m_1)-A{_0}(m_2)+(m_2^2-m_1^2-p^2)B{_0}(p^2,m_1,m_2)}{2p^2}
\end{equation}
with $A_0$ and $B_0$ being the Passarino-Veltman functions.

Because all the superpartners $\phi^i$ have the same mass
$M+M_{soft}$, we can simplify the above four equations to get the final
result of the mass splitting due to SUSY particles :
\begin{equation}
\begin{aligned}
   \Delta M^Q_{SUSY}&=\frac{Q^2}{2}
   (\Sigma_L^{Q}+\Sigma_R^{Q}-\Sigma_L^{0}-\Sigma_R^{0})\\
  &=\frac{g_2^2Q^2}{16\pi^2}[-(V^*_{11}V_{11}+U^*_{11}U_{11})
   B{_1}(\tilde{C_1},\phi)
   -(V^*_{21}V_{21}+U^*_{21}U_{21})B{_1}(\tilde{C_2},\phi)\\
  &+\frac{1}{2}N^*_{i2}N_{i2}B{_1}(\tilde{N_i},\phi)].
\end{aligned}
\end{equation}
The poles in the B1 function are canceled as expected using the unitarity
of the $U, V$ and $N$.  The mass splitting is a function of
$M_1, M_2, \tan\beta, \mu , M_{soft}$ and $M$.  In the correct EW breaking
parameter space, our numerical result for the mass splitting due to SUSY
particles is that
\begin{equation}
   \Delta M^Q_{SUSY} \sim 0.01 Q^2\; {\rm MeV}
\end{equation}
which is negligibly small compared to the pure EW corrections.

\subsection{The thermal relic density}

The thermal relic density fixes the WIMP mass.
In the MDM, the relic species are
$\{\mathrm{X}^{\pm2},\mathrm{X}^{\pm1}, \mathrm{X}^0\}$
and the coannihilation channels are:
\begin{equation}
\mathrm{X}^i\mathrm{X}^j\rightarrow AA, f\bar{f}.
\end{equation}
where $A$ and $f$ denote a EW gauge boson and the SM fermion, respectively.
The mass splittings among them are very small compared to their masses.
In the density calculation, such mass splittings
are negligible. The relic particle thermal average cross section
is\cite{mdm1}:
\begin{equation}
   \langle \sigma_A v\rangle(\mathrm{X}^i\mathrm{X}^j\rightarrow
   \mathrm{AA},f\bar{f})
   \approx\frac{\pi{\alpha}_2^2}{8{M^2}}\times{166} \ ,
\end{equation}
Matching to the relic abundance, the DM particle mass is determined to be
$M=4.4$ TeV without considering Sommerfeld corrections.

Once SUSY is introduced in, a whole bunch of superpartners of those in MDM
will present.  SUSY breaking gives soft masses to the scalars and gauginos.
Looking at $\tilde{\mathrm{X}}$'s, they have a mass $M+M_{soft}$.
In general when $M_{soft}\leq 0.1M$, eg. $M_{soft}\leq 440$
GeV for $M=4.4$ TeV, they have sizable effects on the relic abundance and
must be included in the the relic species \cite{Griest,Gondolo}. So now the
coannihilation relic species are
$\{(\tilde{\mathrm{X}}^{\pm2},\tilde{\mathrm{X}}^{\pm1},
\tilde{\mathrm{X}}^0),(\mathrm{X}^{\pm2},\mathrm{X}^{\pm1},\mathrm{X}^0)\}$
and the coannihilation channels are:
\begin{equation}
\begin{aligned}
  &\tilde{\mathrm{X}}^{i}\tilde{\mathrm{X}}^{j}\rightarrow
   AA, f\bar{f}, \tilde{G}\tilde{G},\\
  &\tilde{\mathrm{X}}^{i}\mathrm{X}^{j}\rightarrow
   f\tilde{\bar{f}}, \tilde{G}A,\\
  &\mathrm{X}^{i}\mathrm{X}^{j}\rightarrow
   AA, f\bar{f}, \tilde{G}\tilde{G},
\end{aligned}
\end{equation}
where $\tilde{G}$ denotes MSSM gauginos and $\tilde{f}$ the superpatner
of $f$.

Furthermore, because $M$ is also much larger than the electroweak scale,
the physics of the above-mentioned coannihilation is basically supersymmetric
and EW gauge symmetric, we can make unbroken EW symmetry and unbroken
supersymmety approximation when calculating the thermal average cross section.
Nevertheless , it is still a hard work. In terms of two-component fields,
there are 24 gauge kinetic vertices and another 24 vertices involving
superpatners. But actually we can have a useful and proper estimate for
the relations between cross sections for the three kinds of processes
in eq. (21). It is found that introducing in SUSY have nearly 4 times
influence on the thermal average cross section. It will be shown in
Fig.2-8 a series of explicit examples by using two-component
spinor techniques\cite{two-component} and the results reads:
\begin{equation}
   v\sigma(\bar{\mathrm{X}}^{+2}\mathrm{X}^{+2}\rightarrow\mathrm{W_3W_3})
   =16\times\frac{8\pi\alpha^2}{3M^2},
\end{equation}
\begin{equation}
   v\sigma(\phi^{*+2}\phi^{+2}\rightarrow\mathrm{W_3W_3})
   =32\times\frac{8\pi\alpha^2}{3M^2},
\end{equation}
\begin{equation}
   v\sigma(\bar{\mathrm{X}}^{+2}\mathrm{X}^{+2}\rightarrow \tilde{\omega_3}\tilde{\omega_3})
   \sim 0,
\end{equation}
\begin{equation}
   v\sigma(\phi^{*+2}\phi^{+2}\rightarrow\tilde{\omega_3}\tilde{\omega_3})
   \sim 0,
\end{equation}
\begin{equation}
   v\sigma(\phi^{+2}\phi^{'-2}\rightarrow\tilde{\omega_3}\tilde{\omega_3})
   \simeq16\times\frac{3\pi\alpha^2}{2M^2},
\end{equation}
\begin{equation}
   v\sigma(\bar{\mathrm{X}}^{+2}\phi^{+2}\rightarrow W_3\tilde{\omega_3})
   \simeq16\times\frac{3\pi \alpha^2}{8M^2},
\end{equation}
\begin{equation}
   v\sigma(\mathrm{X}^{+2}\phi^{'-2}\rightarrow W_3\tilde{\omega_3})
   \simeq16\times\frac{3\pi \alpha^2}{8M^2},
\end{equation}
where $v$ is the relative velocity in the lab frame, Eq. (24),(25) is the
results of the p-wave suppression.

\vspace{10pt}
\begin{figure}[htbp]
\centering
\includegraphics[width=5in]{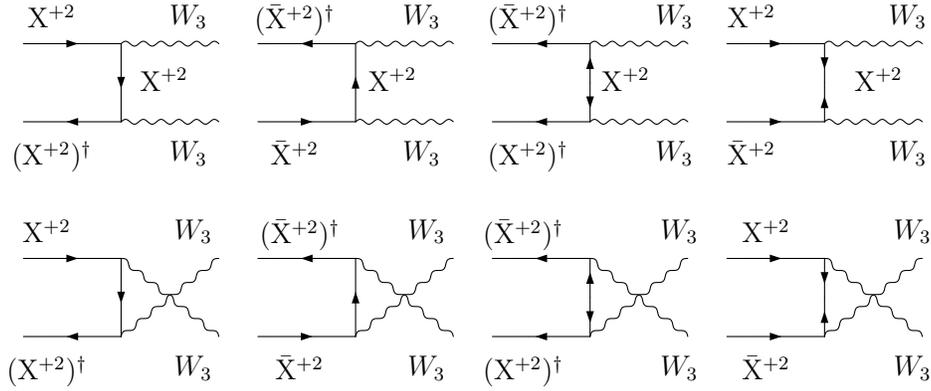}
\caption{The eight Feynman diagrams for $\bar{ X}^{+2}X^{+2}\rightarrow W_3W_3$ }
\end{figure}
\vspace{6pt}
\begin{figure}[htbp]
\centering
\includegraphics[width=4in]{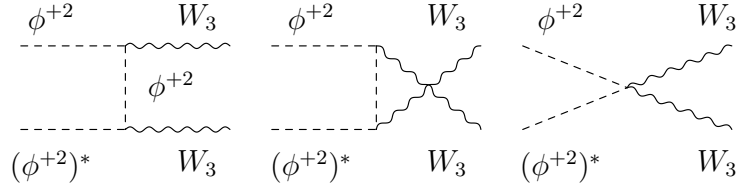}
\caption{The three Feynman diagrams for
$\phi^{*+2}\phi^{+2}\rightarrow\mathrm{W_3W_3}$ }
\end{figure}
\vspace{10pt}
\begin{figure}[!htp]
\centering
\includegraphics[width=5in]{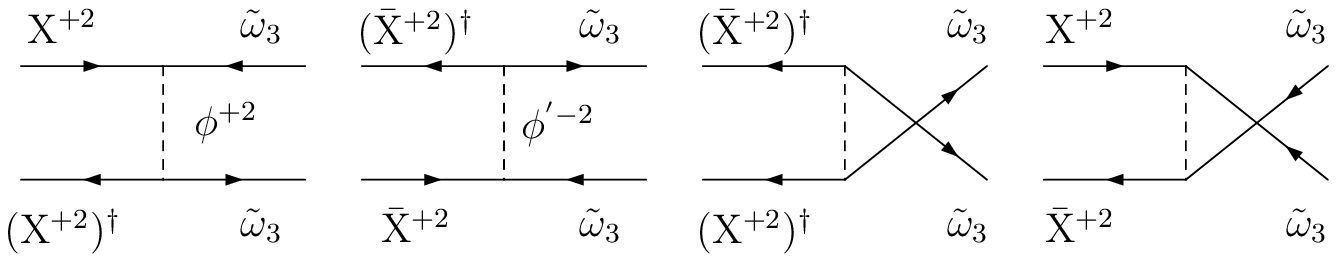}
\caption{The four Feynman diagrams for
$\bar{\mathrm{X}}^{+2}\mathrm{X}^{+2}\rightarrow \tilde{\omega_3}\tilde{\omega_3}$ }
\end{figure}
\vspace{10pt}
\begin{figure}[!htp]
\centering
\includegraphics[width=2.6in]{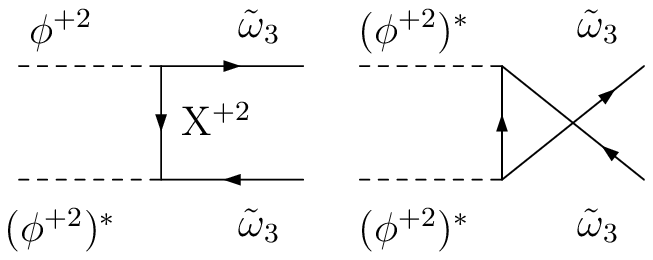}
\caption{The two Feynman diagrams for
$\phi^{*+2}\phi^{+2}\rightarrow\tilde{\omega_3}\tilde{\omega_3}$ }
\end{figure}

\vspace{10pt}
\begin{figure}[!htp]
\centering
\includegraphics[width=5.1in]{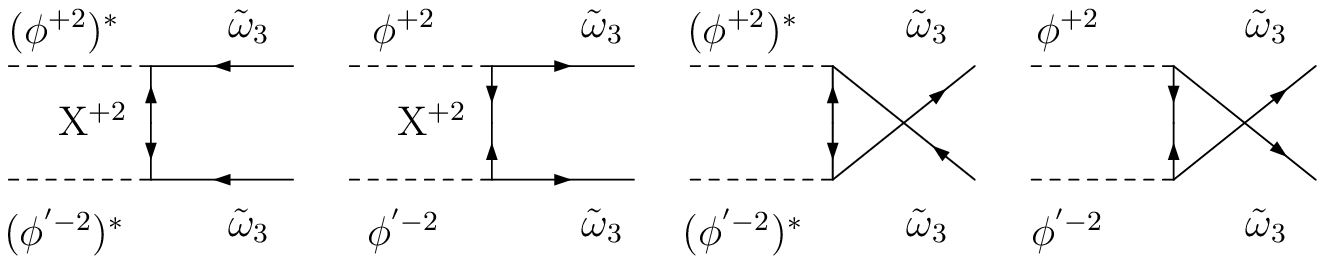}
\caption{The four Feynman diagrams for
$\phi^{+2}\phi^{'-2}\rightarrow\tilde{\omega_3}\tilde{\omega_3}$ }
\end{figure}
\vspace{10pt}
\begin{figure}[!htp]
\centering
\includegraphics[width=5in]{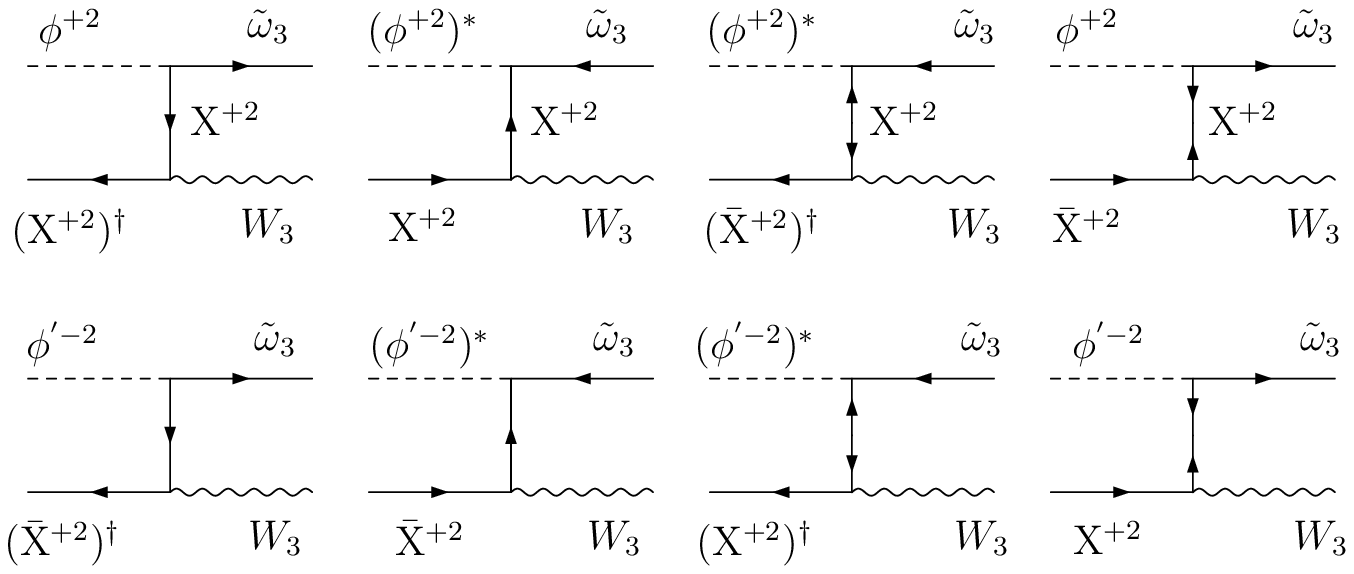}
\includegraphics[width=5in]{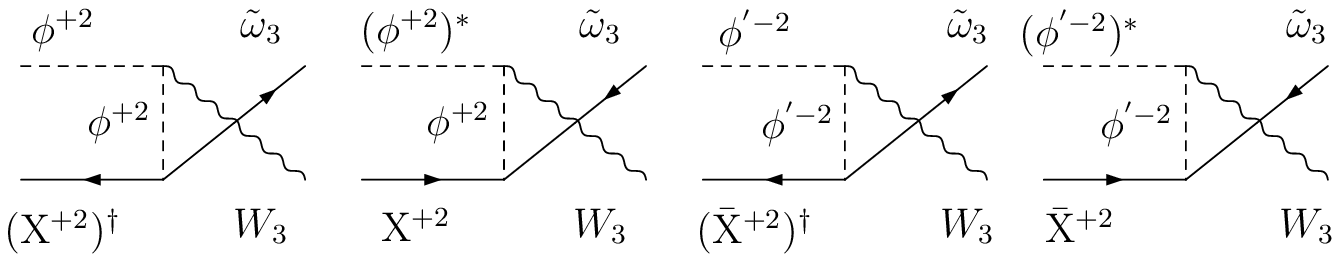}
\caption{The twelve Feynman diagrams for
$\bar{\mathrm{X}}^{+2}\phi^{+2}\rightarrow W_3\tilde{\omega_3}$
and $\mathrm{X}^{+2}\phi^{'-2}\rightarrow W_3\tilde{\omega_3}$ }
\end{figure}

We have already calculate a complete series of thermal cross
section, we end up with:
\begin{equation}
   v\sigma_{eff}=\sum v\sigma\approx
   4v\sigma(\bar{\mathrm{X}}^{+2}\mathrm{X}^{+2}\rightarrow\mathrm{W_3W_3})
\end{equation}
with considering the freedom for $\mathrm{X}^{+2}$, $\bar{\mathrm{X}}^{+2}$
and $\phi^{+2}$,$\phi^{'-2}$ are all $g=2$. This is very different with
the dramatically influence from the higher representation such as $SU$(2)-5
in our problem. Despite we make this conclusion from some specific examples,
we think it is a common result.

Once the DM particle is as heavy as few TeV, the Sommerfeld effect due to
hundreds GeV particles should be taken into consideration.  In the MDM,
the Sommerfeld enhancement effect due to SM particles, especially due to $W$
and $Z$ bosons, has been calculated, the factor is about 5 \cite{sommerfeld}.
We expect approximately the same result in our case.  While our case is
$N = 1$ SUSY, all the SUSY particles should be included in the ladder diagram
calculation in determining the potential which tells us the enhancement
factor.  However including the SUSY particles does not change the Sommerfeld
effect essentially.  In the extreme case of $N = 4$ SUSY, the extra
symmetries just keep the gauge coupling from running.  The potential
itself has the same form as in the non-SUSY Yang-Mills case.   In our $N=1$
SUSY case with soft breaking, the logrithmic runing of the gauge coupling
of the MDM is expected only mildly reduced, the Sommerfeld effect is then
approximatelly the same.  So it is reasonable to say that the Sommerfeld
enhancement factor is about the same as that calculated in the MDM,
$M \simeq \sqrt{5}\times2\times4.4\simeq19.7$ TeV.

\subsection{Direct and Indirect signatures}

As for the direct DM detection rate, this SMDM is of the same order as the
MDM.  The DM particle interacts with quarks via loops.  Although more heavy
particles appear in the loops, the total cross section has the same order,
$\sigma_{\rm SI}\propto 10^{-44}\rm{cm}^2$, which is within the
sensitive of the current experiments, such as Super-CDMS
and Xenon 1-ton\cite{direct1,direct2}.

The DM annihilation in the galaxy may have observable signitures.  The
estimation of the cross section is also like that in the
MDM\cite{signal1,signal2}, the result has no order change.
Note that SUSY does not change the Sommerfeld effect
essentially, the predominant annihilation channel is still into EW
W bosons, $\langle\sigma v\rangle_{WW}\sim10^{-23}{{\rm cm}^3{\rm s}^{-1}}$.
Because the DM mass $M\simeq 19.7$ TeV, we expect this model predict:
(1) continuous rise $e^+ / e^+ + e^-$ spectrum up to about 20 TeV;
(2) flat $ e^+ + e^-$ spectrum up to $M$; (3) $\overline{p}/p$ flux
has excess above the energy probed by PAMELA.
SMDM has the solutions to the PAMELA and Fermi-LAT
anomaly\cite{indirect1,indirect2}, the predictions in higher energies 
need further experimental data to verify.

\section{Summary and Discussion}

We have made SUSY extension to the MDM model by introducing a complex
scalar quintuplet as the superpartner of the fermion quintuplet.  The
neutral component of the fermionic 5-plet is still the DM particle as in
the MDM model.  Mass splittings among the fermionic 5-plets due to SUSY
have been calculated in detail, they are found to be small.  By
considering new relic species and new coannihilation channels into the
MSSM final states, the DM mass is estimated to be 19.7 TeV.

The direct and indirect signals are basically the same as those in the MDM.
Numerically the DM elastic scattering cross section with a nucleus is about
$10^{-44}$ cm$^3$ s$^{-1}$ and the cross section of the predominant
annihilation channel into W bosons is about
$10^{-23}$ cm$^3$ s$^{-1}$. SMDM predicts $e^+, e^++e^-, \overline{p}$
spectrum in agreement with the previous PAMELA, Fermi-LAT data,
$\overline{p}$ flux has excess above the energy probed by PAMELA
which need further experimental test.

In the near future, suppose SUSY is discovered, say at LHC, it will be
still a question if MSSM itself provides a DM particle, because R-parity
as a discrete symmetry is still an assumption which is not as solid as
gauge invariance and SUSY.  It is plausible that R-parity is violated.
In that case, it is still simple and interesting to have the DM via
introducing SU(2)$_L$ high dimensional representations.

\acknowledgments
We would like to thank Yu-Qi Chen, Xiao-Jun Bi, Jia-Shu Lu and Hua Shao
for helpful discussions. This work was supported in part by the National
Natural Science Foundation of China under Nos. 11075193 and 10821504,
and by the National Basic Research Program of China under
Grant No. 2010CB833000.

\appendix*

\section{the su(2)-n representation}

The SU(2)-n representation $U^j$ ( n=2j+1 ) is self-conjugate , when $j$
is integer ($n$-odd), $U^j$ is real , when $j$ is half-integer
($n$-even), $U^j$ is self-conjugate also but not real :
\begin{equation}
   X U^j X^{-1}= {U^j}^*\Rightarrow
\begin{cases}
   U^j={U^j}^* & j=0,1,2,...\ ,\  X \ \rm{is\ symmetric} \\
   U^j \ \rm{is\ not\ real}&j
   =\frac{1}{2},\frac{3}{2},...\ ,\ X \ \rm{is\ antisymmetric}
\end{cases}
\end{equation}
it is important in proving some identities including the SU(2)
generators.

The generators of the su(2)-n representation is:
\begin{equation}
\begin{aligned}
  &(T_1^j)_{\nu\mu}=\frac{1}{2}
   [\delta_{\nu(\mu+1)}\Gamma_\nu^j+\delta_{\nu(\mu-1)}\Gamma_{-\nu}^j]\\
  &(T_2^j)_{\nu\mu}=-\frac{i}{2}
   [\delta_{\nu(\mu+1)}\Gamma_\nu^j-\delta_{\nu(\mu-1)}\Gamma_{-\nu}^j]&\\
  &(T_3^j)_{\nu\mu}=\mu\delta_{\nu\mu} .
\end{aligned}
\end{equation}
where
\begin{equation}
   \Gamma_\nu^j=\Gamma_{-\nu+1}^j={(j+\nu)(j-\nu+1)}^{1/2}
\end{equation}
for example
\begin{equation}
T_3^1=\frac{\sqrt{2}}{2}
\begin{pmatrix}
0 & 1 & 0
\\1 & 0 & 1
\\0 & 1 & 0
\end{pmatrix},
T_3^2=\frac{\sqrt{2}}{2}
\begin{pmatrix}
0 & -i & 0
\\i & 0 & -i
\\0 & i & 0
\end{pmatrix},
T_3^3=
\begin{pmatrix}
1 & 0 & 0
\\0 & 0 & 0
\\0 & 0 & -1
\end{pmatrix},
\end{equation}

\begin{equation}
T_5^1=
\begin{pmatrix}
0 & 1 & 0 & 0& 0
\\1 & 0 & \frac{\sqrt{6}}{2} & 0 & 0
\\0 & \frac{\sqrt{6}}{2} & 0 & \frac{\sqrt{6}}{2} & 0
\\0 & 0 & \frac{\sqrt{6}}{2} & 0 & 1
\\0 & 0 & 0 & 1 & 0
\end{pmatrix},
T_5^2=
\begin{pmatrix}
0 & -i & 0 & 0 & 0
\\i & 0 & -i\frac{\sqrt{6}}{2} & 0 & 0
\\0 & i\frac{\sqrt{6}}{2} & 0 & -i\frac{\sqrt{6}}{2} & 0
\\0 & 0 & i\frac{\sqrt{6}}{2} & 0 & -i
\\0 & 0 & 0 & i & 0
\end{pmatrix},
T_5^3=
\begin{pmatrix}
2 & 0 & 0 & 0 & 0
\\0 & 1 & 0 & 0 & 0
\\0 & 0 & 0 & 0 & 0
\\0 & 0 & 0 & -1 & 0
\\0 & 0 & 0 & 0 & -2
\end{pmatrix}.
\end{equation}
and have
\begin{equation}
\bar{T^a}T^2=-T^2T^a
\end{equation}



\begin{thebibliography}{99}

\bibitem{mdm1}
M. Cirelli, N. Fornengo, A. Strumia,
Nucl.Phys. B753 (2006)
arXiv:hep-ph/0512090.

\bibitem{mdm2}
M. Cirelli and A. Strumia,
New J. Phys. 11 (2009) 105005 arXiv:0903.3381.

\bibitem{others}
Y. Cai, X.-G. He, M. Ramsey-Musolf, and L.-H. Tsai, arXiv:1108.0969;\\
C.-H. Chen and S.S.C. Law, arXiv:1111.5462.

\bibitem{susy}
For reviews, see
M. E. Peskin, arXiv:0801.1928;
S. P. Martin, arXiv:hep-ph/9709356.

\bibitem{LSP}
G. Bertone, D. Hooper and J. Silk,
Phys. Rep. {\bf 405}, 279-390, (2005).

\bibitem{Griest}
G. Jungman and M. Kamionkowski and K. Griest, Phys. Rep. {\bf 267}, 195 (1996).

\bibitem{Gondolo}
J. Edsjo and P. Gondolo, \PRD56,1879-1894,(1997).

\bibitem{two-component}
H. K. Dreiner, H. E. Haber and S. P. Martin, arXiv:0812.1594.

\bibitem{sommerfeld}
M. Cirelli, A. Strumia, Matteo Tamburini,
Nucl.Phys. B787 (2007), arXiv:0706.4071.

\bibitem{direct1}
Z.Ahmed {\it et al.}  [CDMS Collaboration], Phys. Rev. Lett. 102, 011301 (2009).

\bibitem{direct2}
J.Angle {\it et al.}  [XENON Collaboration], arXiv: 1104.2549.

\bibitem{signal1}
M. Cirelli and A. Strumia,
PoS IDM 2008 (2008) 089 arXiv: 0808.3867.

\bibitem{signal2}
M. Cirelli, R. Franceschini, A. Strumia
Nucl. Phys. B 800:204-220,2008
arXiv:0802.3378.

\bibitem{indirect1}
O. Adriani {\it et al.} [PAMELA Collaboration], arXiv:1103.2880.

\bibitem{indirect2}
A. A. Abdo {\it et al.} [Fermi LAT Collaboration],
Phys. Rev. Lett. 102 (2009).




\end{thebibliography}
\end{document}